\documentclass[aps,prl,groupedaddress,twocolumn]{revtex4-2}

\usepackage{amsmath}
\usepackage{mathrsfs}
\usepackage{ulem}
\usepackage{graphicx}
\usepackage{xcolor}

\def\d{{\rm d}}

\def\D{\mathscr{D}}
\def\AT{\tilde{A}}

\begin{document}


\title{Chaotic synchronization in optical frequency combs}

\author{G\"unter Steinmeyer}
\email[Corresponding author: ]{steinmey@mbi-berlin.de}
\affiliation{Max Born Institute for Nonlinear Optics and Short Pulse Spectroscopy, Max-Born-Stra\ss e 2a, 12489 Berlin, Germany}
\affiliation{Institut f\"ur Physik, Humboldt Universit\"at zu Berlin, Newtonstra\ss e 15, 12489 Berlin, Germany}

\author{Weidong Chen}
\affiliation{Max Born Institute for Nonlinear Optics and Short Pulse Spectroscopy, Max-Born-Stra\ss e 2a, 12489 Berlin, Germany}
\affiliation{Fujian Institute of Research on the Structure of Matter, Chinese Academy of Sciences, 350002 Fuzhou, China}





\date{\today}

\begin{abstract}
Using a discrete mode approach we investigate the intermodal dynamics in a frequency comb with nonlinear coupling due to four-wave mixing. In the presence of sufficient saturable absorption, phase space collapses into a single state, and all modes tightly lock with identical phase. In case of a purely reactive nonlinearity, a less constrained locking mechanism emerges, which keeps phase differences of neighboring modes loosely bounded and leads to the formation of quasi-periodic breathers. Analysis of the underlying nonlinear correlations identifies a chaotic synchronization process as the origin of this previously unreported locking mechanism. Despite their rather large Lyapunov exponent and correlation dimension, the highly chaotic dynamics may be easily overlooked by common diagnostic approaches for laser mode-locking. This finding may finally explain numerous unexplained reports on laser self mode-locking in the absence of an effective saturable absorber mechanism and the formation of self-frequency modulated combs in semiconductor lasers.
\end{abstract}


\maketitle

Synchronization between independent oscillators is a ubiquitous phenomenon in physics, which was first described by Christiaan Huygens in the 17th century \cite{synch,Huygens}. Mounting two clocks with nearly identical frequencies on a common base, Huygens observed that these clocks would synchronize in antiphase. Analysis reveals that movements of the common base induce a nonlinear coupling mechanism between the phases of the two clock pendulums \cite{Metronome}. Similar synchronization mechanisms have been observed in a large range of physical \cite{Plasma,CO2Laser}, chemical \cite{Chem,Kuramoto} and electronic systems \cite{vanderPol}. Synchronization also plays a role in biological systems \cite{Firefly,Neuron}, e.g., coupling the main rhythmic processes in the human cardiovascular system \cite{Cardio}. Large-scale synchronization of myocites in the sinoatrial node acts as a natural pacemaker in the human heart. If this synchronization is hampered, the electrical signal from the sinotrial node may become chaotic, resulting in the life-threatening condition of atrial fibrillation.

In the following, we investigate large-scale synchronization effects in a much simpler system of nonlinearly coupled optical oscillators. In such a multimode laser, the longitudinal modes of the laser cavity naturally form a comb of nearly equidistantly spaced oscillators with frequencies $f_i$ \cite{Siegman}. Dispersion of intracavity materials causes a deviation from perfect equidistance, similar to the slight detuning of the clocks in Huygens's experiment. Nonlinear optical effects may take the role of the common base and couple individual modes \cite{Mandel}. Using a sufficiently strong dissipative nonlinearity, all oscillators synchronize, and a perfect frequency comb with $f_i=f_0+i \Delta f$ results \cite{Haus1,Haus2,Haus3}. This process is known as mode-locking and leads to femtosecond pulse trains, which have found a plethora of applications in ultrafast spectroscopy and precision frequency metrology. In contrast to this well understood mechansim, self-mode-locking and self comb formation have also been observed in the clear absence of the dissipative nonlinearity \cite{Rikken,SESAMfree,negativeKerr,Koch,Bimberg,Threshold,Faist} necessary to stabilize the mode-locking process \cite{Haus2}. This self mode-locking process was often explained with the presence of a strong four-wave mixing nonlinearity, and the resulting combs have been termed as self-frequency modulated (FM) combs.

\begin{figure}[tb]
\begin{center}
\includegraphics[width=0.7 \linewidth]{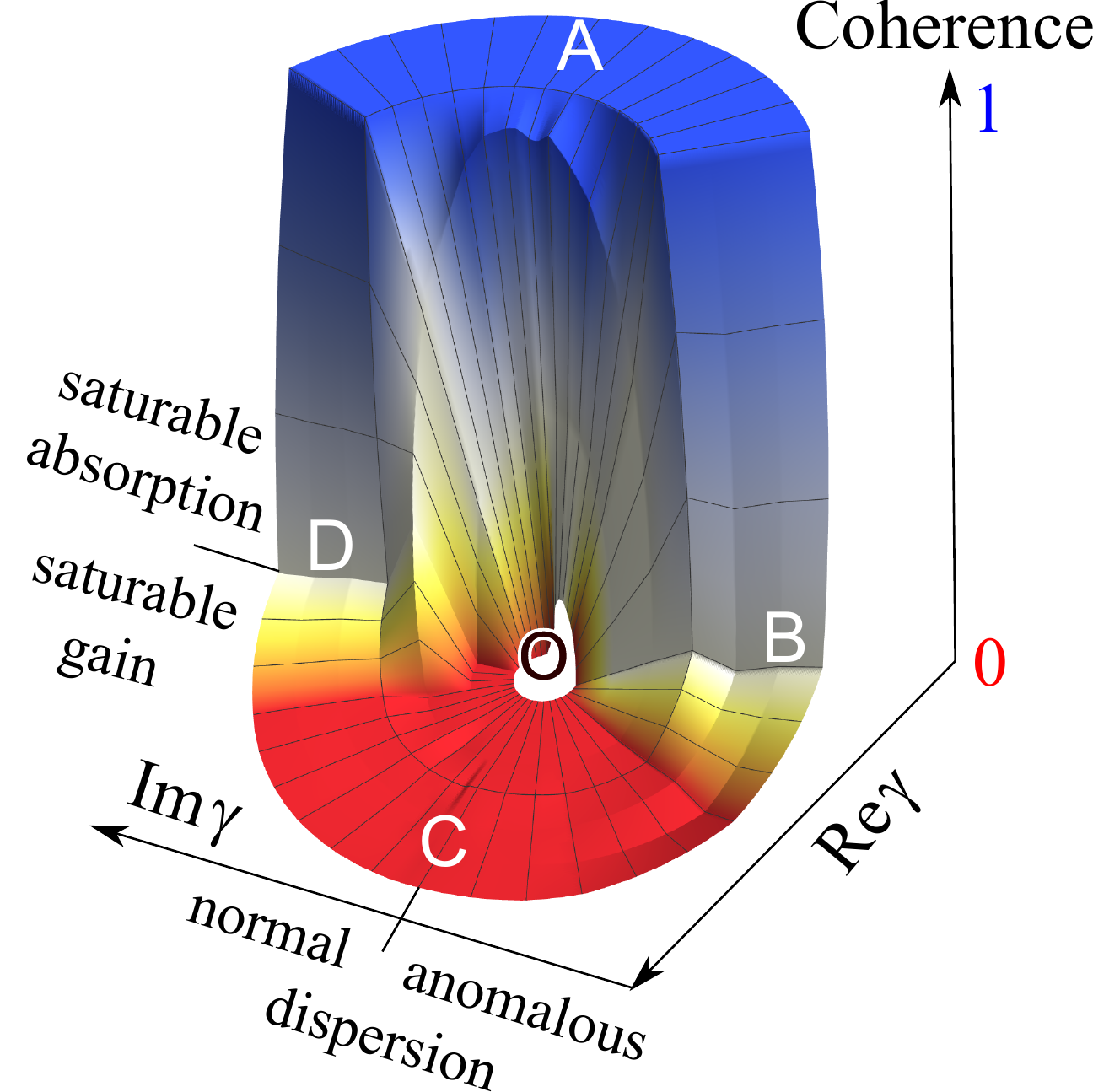}
\end{center}
\caption{Coherence of a multimode laser in the presence of a four-wave mixing nonlinearity. $\gamma$ parametrizes the ratio of reactive nonlinearity and second-order dispersion, $\delta$ a dissipative nonlinearity. Point A indicates the perfect coherence for dominant saturable absorption\cite{Haus3}, whereas O and C refers to the situation of phase turbulence \cite{CGLE}. Points B and D indicate frequency-locking due to a purely reactive nonlinearity in the presence of anomalous and normal dispersion, respectively.}
\label{fig:coherence}
\end{figure}

\begin{figure}[tb]
\begin{center}
\includegraphics[width=0.95 \linewidth]{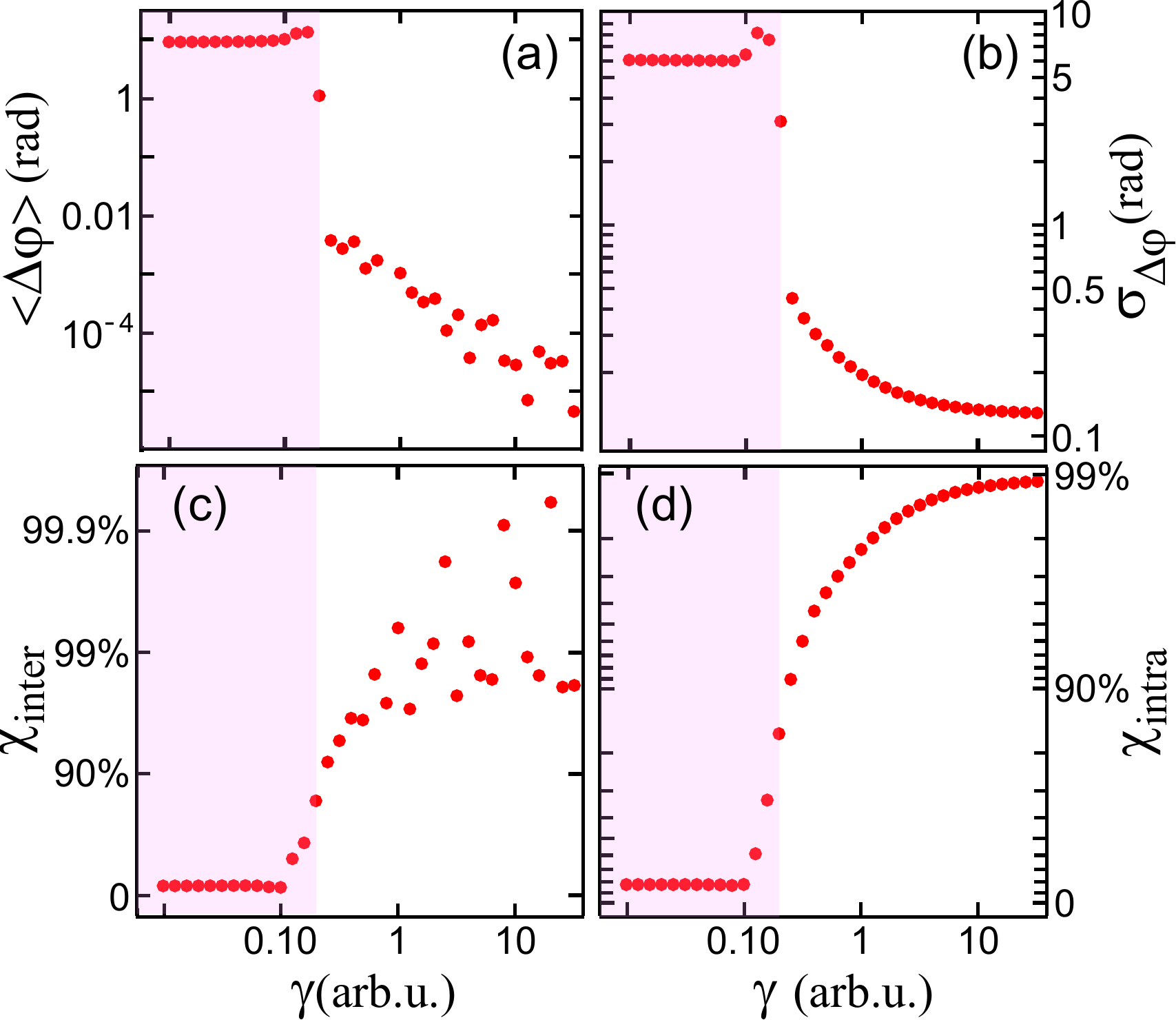}
\end{center}
\caption{Dependence of phase variability and coherence on the strength of a purely reactive nonlinearity $\gamma$. $\delta=0$, $\D=1$. (a) Average value of the phase differences between adjacent modes $<\Delta\varphi>$. Exceeding a threshold at $\gamma\approx 0.2$, phase differences collapse to sub-mrad values, which results in a concomitant reduction of the beat note width. (b) Fluctuations of the phases $\sigma_{\Delta\varphi}$. (c) Interpulse coherence [Eq.~(\ref{eq:inter})]. (c) Intrapulse coherence [Eq.~(\ref{eq:intra})]. Red shading indicates region of phase turbulence \cite{CGLE}.}
\label{fig:cohscan}
\end{figure}

We analyze synchronization effects in multimode lasers in the presence of an arbitrary third-order nonlinearity. Our approach is based on the Haus master equation, which has enabled theoretical understanding of all mode-locking variants to date, with the noted exception of self mode-locking \cite{Haus1,Haus2,Haus3}. This equation is a variant of the complex Ginzburg-Landau equation (CGLE) \cite{CGLE}
\begin{eqnarray}
\partial_z A(z,t) & = & i \gamma |A|^2 A + i \D \partial_t^2 A \nonumber \\
 & & - \delta |A|^2 A - g (q-q_0) A, \label{eq:Haus}
\end{eqnarray}
in which we added a gain saturation term \cite{Haus1} to warrant energy conservation in the system. This partial differential equation describes the propagation of the complex-valued electric-field envelope $A(t)$ along the propagation coordinate $z$ and includes effects of group-velocity dispersion $\D \partial_t^2 A$, the instantaneous Kerr nonlinearity $\gamma |A|^2 A$, and fast saturable absorption $\delta |A|^2 A$. Slow gain saturation in included via the final term in Eq.~(\ref{eq:Haus}). Here $q_0$ is the laser pulse energy $q$ under steady-state conditions, and negative feedback is provided if the gain $g q_0 A$ does not equal losses $g q A$. Usually, the Haus master equation is treated as an eigenvalue problem of the differential operator on the rhs of Eq.~(\ref{eq:Haus}), i.e., one seeks soliton solutions that obey $\partial_z A = 0$. However, the existence of solitons is only a necessary and not a sufficient criterion for mode-locking \cite{Haus2}. Dropping all dissipative terms, e.g., Eq.~(\ref{eq:Haus}) becomes the nonlinear Schr\"odinger equation with its hyperbolic secant soliton, which is, however, unstable without a stabilizing dissipative term $\delta>0$ in Eq.~(\ref{eq:Haus}). In order to obtain access of the dynamics of the individual modes, we study the master equation in the frequency domain \cite{Lamb,Brunner,Cagesolitons}. The partial differential Eq.~(\ref{eq:Haus}) now becomes a set ($j=-n,\dots,+n$) of coupled ordinary differential equations
\begin{equation}
\partial_z \AT_j(z) = (i \gamma - \delta) \!\!\!\!\! \sum_{j=k+\ell-m} \!\!\!\!\! \AT_k \AT_\ell \AT_m^* + \left[i \beta_j - g_j (q-q_0)\right] \AT_j, \label{eq:master}
\end{equation}
which is related to the Kuramoto model of multimode lasers \cite{Cluster,Kuramoto}.
Temporal derivatives $\D \partial_t^2 A$ are now replaced by
$\beta_j = \D j^2 \Delta\omega^2 /2$ with the mode spacing $\Delta\omega$, and the nonlinearities are expressed as convolution sums. While the number of coupling terms grows with the third power of $n$, Eq.~(\ref{eq:master}) can still be integrated out with high accuracy ($|q-q_0|<10^{-5} q_0$) and at reasonable computation times for $n<80$, using an Adams predictor-corrector method. In the following, we judge the coherence of the laser from computing the temporal or interpulse coherence
\begin{equation}
\chi_{\rm inter} (\Delta z) = \int_0^Z  \AT_k(z) \AT_k^*(z+\Delta z) \d z . \label{eq:inter}
\end{equation}
Additionally, we can also evaluate the spectral or intrapulse coherence \cite{Intrapulse} from
\begin{equation}
\chi_{\rm intra} (k,\ell) = \int_0^Z  \AT_k(z) \AT_\ell^*(z) \d z. \label{eq:intra}
\end{equation}
Varying the complex sign of the nonlinearity $-\delta+i \gamma$ in Eq.~(\ref{eq:master}), we exclusively find stable mode-locking with perfect coherence for positive values of $\delta$, i.e., saturable absorption and ratios $\gamma/\beta>1$ (region A in Fig.~\ref{fig:coherence}). In contrast, regions with negative $\delta$ exhibit near-vanishing values of $\chi_{\rm inter}$ and $\chi_{\rm intra}$ (region C in Fig.~\ref{fig:coherence}). This confirms the well-known paradigm that saturable gain cannot stabilize mode-locking \cite{Haus2} and leads to phase turbulence \cite{CGLE}.
Apart from these well understood regions, however, we find a regime of partial coherence $0.9 < \chi < 1$ for the case of a vanishing dissipative nonlinearity $\delta$ and sufficiently large reactive nonlinearity $|\gamma/\beta|>0.2$, see Fig.~\ref{fig:cohscan}(c,d) and regions B and D in Fig.~\ref{fig:coherence}. In this regime, the phase differences $\Delta\varphi_{k\ell}$ between individual modes $i$ and $j$ may show large temporal variations around a constant average value. The phase dynamics nevertheless remain bounded at values $|\Delta\varphi{k(k+1)}|<\pi/4$, i.e., the four-wave mixing nonlinearity effectively induces a frequency lock between the modes rather than the tight phase lock characteristic for traditional mode-locking \cite{Siegman}. The former behavior is typical for chaotic synchronization between nonlinear oscillators \cite{Bounded}.

The threshold-like transitions \cite{Threshold} between these regimes can be probably best understood in the analogy of phase transitions. In case of insufficient nonlinearity, the modes evolve independently, similar to the movement of molecules in a gas. If present, saturable absorption induces a tight bond between the modes, which consequently condense in a soliton state with periodic waveform. In between these two extreme states, there exists a cohesion-like interaction mechanism, which still allows for phase variations between neighboring modes. This regime can therefore also be understood as the liquid state of mode-locking.

\begin{figure}[tb]
\begin{center}
\includegraphics[width=0.95 \linewidth]{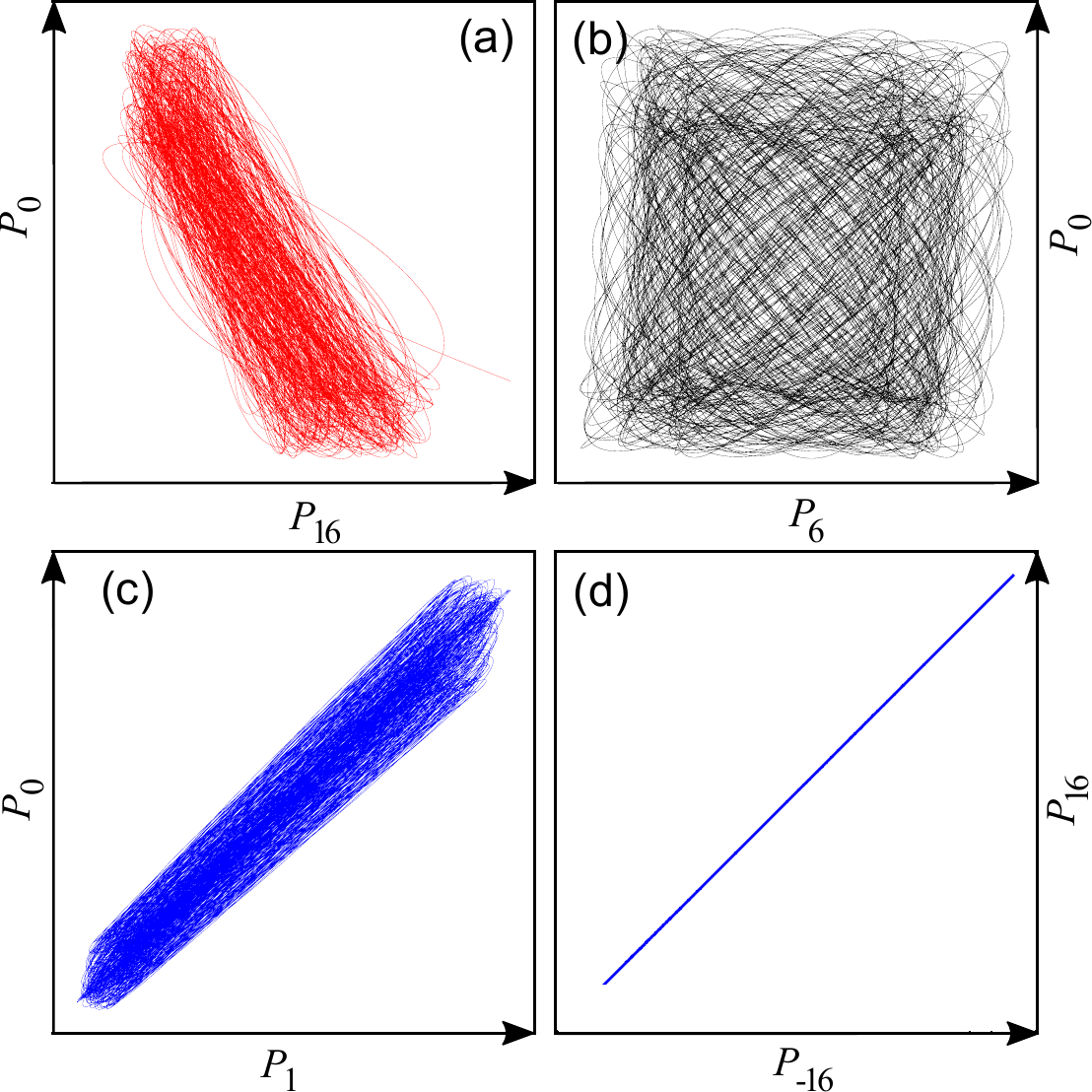}
\end{center}
\caption{Parametric plot of the power contents in center and off-center modes based on solution of Eq.~(\ref{eq:master}) with $n=16$. Phase space trajectories have been plotted for a large number ($\approx 400$) of quasi-periodic cycles to show the absence of ergodicity and the relatively low entropy of the system compared to phase turbulence. (a) Center mode power $P_0$ vs. $P_{16}$. (b) $P_0$ vs. $P_{6}$. (c) $P_0$ vs. adjacent $P_1$. $P_{16}$ vs. $P_{-16}$. }
\label{fig:corr}
\end{figure}

In order to understand the physical mechanism behind the chaotic synchronization, we integrated out a 31-mode version of Eq.~(\ref{eq:master}) and plotted recurrence plots between the power in the central mode $P_0=|\AT_0|^2$ and the power of modes at increasing spectral separation $P_1, P_6,$ and $P_{16}$. As can be seen from Fig.~\ref{fig:corr}(a) and (c), there is a strong yet not perfect correlation between $P_0$ and $P_1$ and a similar anticorrelation between $P_0$ and $P_{16}$. In contrast, there is vanishing correlation between $P_0$ and $P_6$ [Fig.~\ref{fig:corr}(b)]. This anticorrelation phenomenon leads to mode partition noise and manifests itself as a quasi-periodic, chaotic energy exchange between spectral wings and center modes \cite{Mandel,Cluster} while the spectrally integrated laser power may remain perfectly constant. Previous experimental observations indicate typical spectral breathing frequencies in the range from 10 to several hundred MHz \cite{Mandel,Liang}. Given the neglection of higher-order dispersion in our model equations, the four-wave mixing process induces a tight phase lock between any two opposite modes with equal separation to the center mode [Fig.~\ref{fig:corr}(d)]. The quasi-periodic character of the breathing becomes clear from Fourier-transforming $P_0(t=z/c)$. Resulting spectra exhibit one dominant breathing frequency with a number of incommensurable sidebands. The main breathing frequency $f_0(t)$ shows large and rapid variations, which are again an indication for the presence of a chaotic synchronization process \cite{Bounded}.

For a verification of the chaotic nature of the dynamics, we computed correlation sums from a large data set (25,000 samples) and determined the largest Lyapunov exponent $\lambda$, the correlation dimension $D$ as well as the information entropy $S$ \cite{Analysis}. As we have complete access to the phase space of our system, we can rather directly determine these characteristics of chaotic behavior without having to reconstruct the phase space from a one-dimensional data set first. For this analysis we deliberately chose a condition $|\gamma/\beta|=4$, which leads to near-unity values of coherence $\chi_{\rm inter} \approx 98\%$. Results are summarized in Table \ref{tab:results}.
\begin{table}[b!]
\begin{tabular}{ c || c | c | c | c || l}
Regime & $\chi_{\rm inter}$ & $S$  & $D$ & $\lambda$ & \\
\hline
A & 1 & 0 & 0 & 0 & regular \\
B & 0.978 & $43\%$ & $5 \pm 1$ & 3.2 $\overline{f_0}$ & chaotic QP\\
D & 0.985 & $23\%$ & $1.5 \pm 0.1$ & 2.3 $\overline{f_0}$ & chaotic QP \\
O & 0 & $100\%$ & 20 & - & turbulent
\end{tabular}
\caption{Characteristics of the nonlinear dynamics of the quasi-periodic behavior of the system. Letters in first column correspond to labels in Fig.~\ref{fig:coherence}. $\chi_{\rm inter}$: interpulse coherence. $S$: information entropy, normalized to case C. $D$: correlation dimension as an estimate for phase space dimension. $\lambda$: Lyapunov exponent per average period of the quasi-periodicity (QP).} \label{tab:results}
\end{table}
For comparison, we included the respective characteristics for regular mode-locking and phase turbulence if applicable. Condensation in a soliton state, e.g., results in the collapse of phase space to a single state [point A in Fig.~\ref{fig:coherence}] . Correspondingly, the entropy $S$ vanishes and $D=0$ . In the absence of any coupling mechanism, phase turbulence fills the entire available phase space [point O in Fig.~\ref{fig:coherence}], and $S=1$. $D$ assumes the highest possible value $D=20$ in this case. Inspecting the values for $D$ and $S$ for the purely conservative regimes B and D, one observes intermediate values for both characteristics, as would be expected in a liquid state of mode-locking, where the quasi-periodic oscillation leads to a reduction of phase space dimension and the system is non-ergodic. It can also be clearly seen that regime B (focusing Kerr nonlinearity and anomalous dispersion) is apparently more chaotic than regime D (normal dispersion). This is readily understood from the presence of a modulation instability for anomalous dispersion. This conclusion is also supported by the analysis of the Lyapunov exponent of scenarios B and D. In the anomalous dispersion regime, the phase space trajectories have already diverged by a factor $\exp(-1)$ after a third of an average breathing period $\tau_B=\overline{f_0}^{-1}$; for normal dispersion this happens after half a cycle.

\begin{figure}[t]
\begin{center}
\includegraphics[width=0.95 \linewidth]{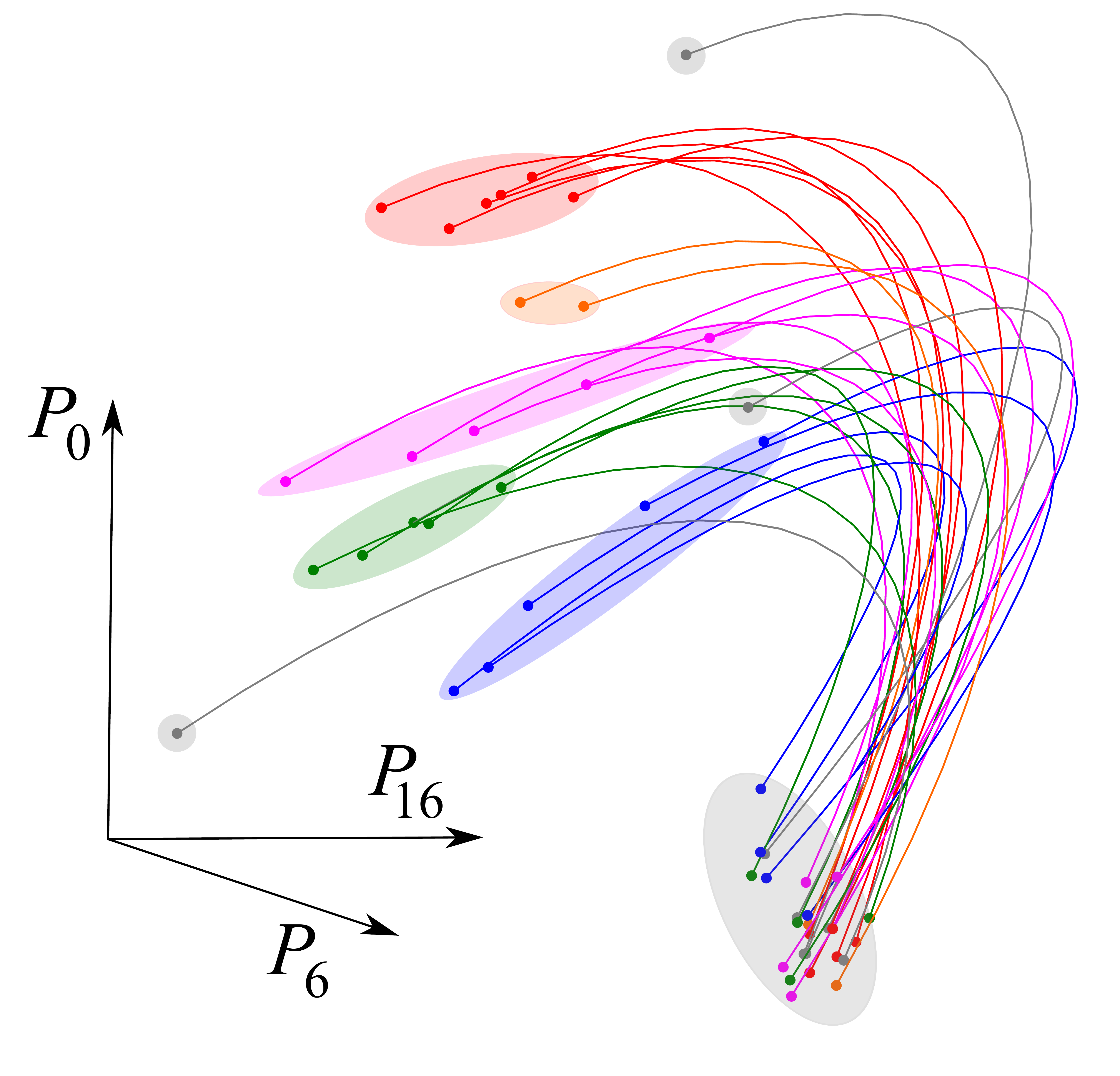}
\end{center}
\caption{Visualization of the chaotic phase space dynamics. 3-dimensional sub-space consisting of $P_0$,$P_6$,$P_{16}$. Parameters as in Fig.~\ref{fig:corr}. Gray ellipse: set of 25 closest neighbors within a long data set comprising $\approx 400$ quasi-periodic cycles. Colored ellipses: concentration points of phase space trajectories 0.25 cycles later.}
\label{fig:lyap}
\end{figure}

For a better illustration of the underlying chaotic dynamics in this quasi-periodic system, we selected a short ($0.1 \tau_B$)  sequence out of a long time series that we obtained from integrating Eq.~(\ref{eq:master}). We then searched for the 25 nearest neighbors and plotted their phase space trajectories in a three-dimensional subspace, see Fig.~\ref{fig:lyap}. To this end, we extracted the power content in the central mode and two modes in the spectral wings of the breather
 ${P_0(t), P_6(t),P_{16}(t)}$. The time span shown ($0.25 \tau_B$) was chosen close to the inverse Lyapunov exponent $\lambda^{-1}$ of the system. As Fig.~\ref{fig:lyap} shows, the phase space trajectories diverge quickly on the time scale shown and regroup at about 4 or 5 new positions in phase space (highlighted in different colors). If one continues to track these newly grouped trajectories for another $\lambda^{-1}$, one finds that they continue to diverge in a fractal fashion. As four-wave mixing induces a pronounced anti-correlation between spectral wings and center modes, the system is not ergodic as in the turbulent regime, but is confined to a $D \approx 5$ dimensional subset of the total available phase space.

\begin{figure}[tb]
\begin{center}
\includegraphics[width=0.95 \linewidth]{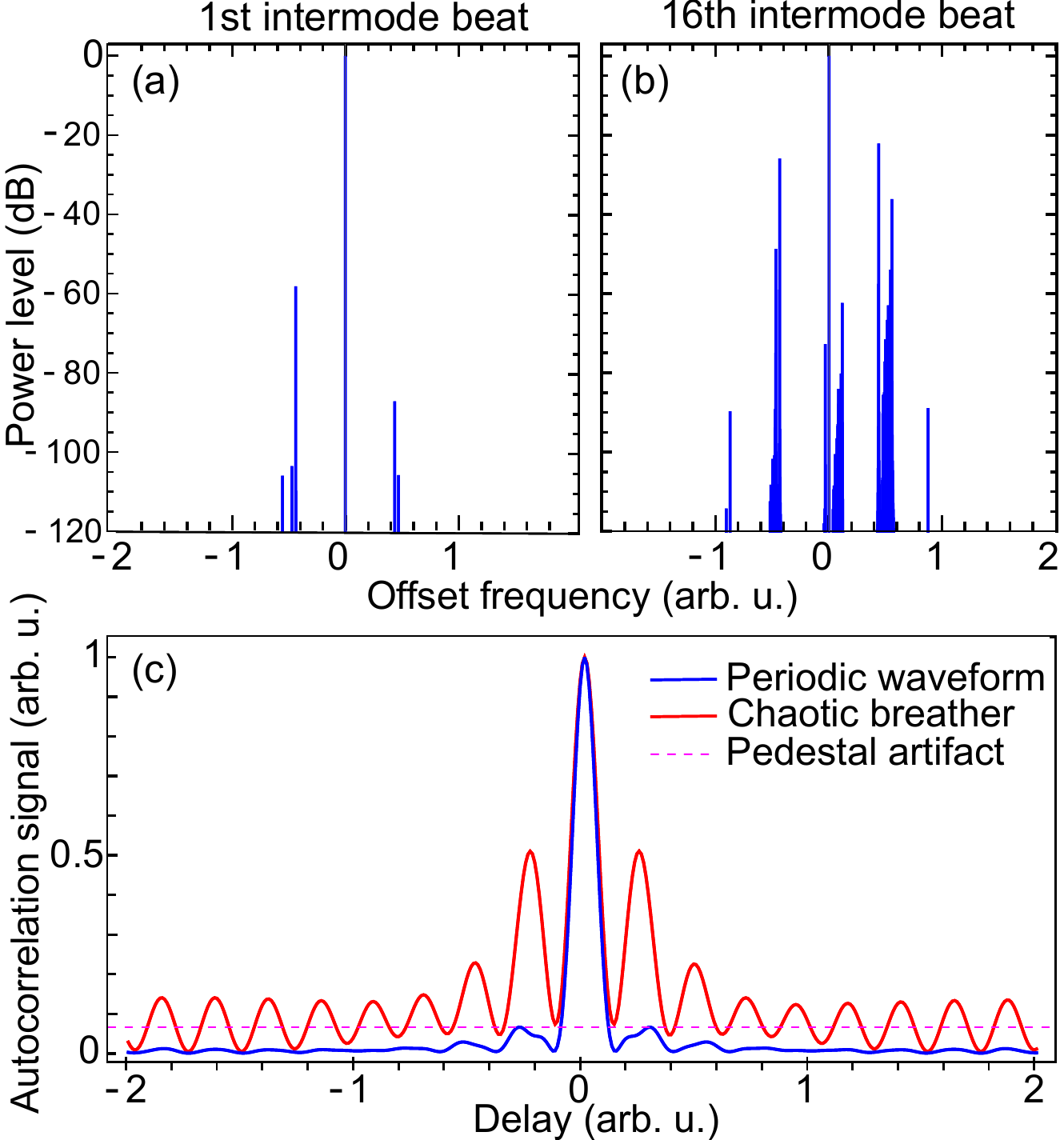}
\end{center}
\caption{Simulated laser diagnostics. $n=16$ (a) First intermode beat note. (b) 16th intermode beat note. (c) Intensity autocorrelations $g^{(2)}(\tau)$. Red curve: chaotic breather. Blue curve: Periodic waveform with identical Fourier limit. Dashed line: Pedestal artifact of FM mode-locking.    }
\label{fig:beat}
\end{figure}

Given the highly chaotic dynamics of the modes that are coupled by a pure reactive nonlinearity, the question arises why this highly chaotic nature has been overlooked in most of the previous literature. In the mode-locked laser community, one typically measures autocorrelations and radio frequency spectra of the individual beat tones \cite{Tropper}. We therefore simulated such measurements. As shown in Fig.~\ref{fig:beat}(a), the fundamental beatnote may appear as a near perfect narrow line with indications of the underlying breather dynamics at a power level of $-60\,$dB, which is easily overlooked in laser characterization. In contrast, going to a higher beat note [Fig.~\ref{fig:beat}(b)], the quasi-periodic breather dynamics are at a measurable level. However, such high beat notes are often not accessible in semiconductor lasers with their multi-GHz mode spacing. A similar problem arises in autocorrelation, which normally indicate a degraded coherence by the presence of a pedestal artifact \cite{Walmsley,Ratner,Pedestal}. In the absence of interpulse coherence, pedestal and central coherence spike exhibit a ratio of 1:1. In Fig.~\ref{fig:beat}(c), the ratio is about 1:10, and the pedestal may further decrease if the nonlinear coupling in the system is increased. While correlation-based methods measure a central artifact that corresponds to the autocorrelation of the transform-limited pulse shape, we find that spectral interferometry based methods \cite{Standards} indicate the presence of a strongly chirped pulse. Varying the coupling over a large parameter range, we observe that traditional characterization methods only allow for a detection of the partial coherence for near-unity ratios of $\beta/\gamma$ in Eq.~(\ref{eq:master}). In the latter case, the phase excursions of the individual modes are on the order of 0.5\,rad, and the pulses cannot be compressed anywhere close to the Fourier limit. Currently, the only way to safeguard against a coherence degradation in an FM comb appears to be a spectral separation of beat notes with a monochromator \cite{Mandel}.

In conclusion, we discussed a new regime of mode-locking that may arise in the perfect absence of saturable absorption and even in the presence of weak gain saturation effects. This regime can be understood as the liquid state of mode-locking, offering a bounded phase rather than a static phase lock between adjacent modes, similar to molecular movement in a liquid. We believe that this previously poorly understood mechanism explains a majority of the observations of self mode-locking and self formation of frequency-modulated combs in semiconductor lasers \cite{Koch,negativeKerr,Bimberg,Threshold,Faist,SESAMfree,Liang}. Within this analogy, one observes a phase transition at the onset of the locking effect, which leads to a threshold-like collapse of the beat note width \cite{Threshold} and to formation of a frequency comb. Despite the apparent comb structure, the laser may operate in a perfectly continuous way, with no discernible power fluctuations in the spectrally integrated signal \cite{Mandel}. The underlying chaotic, quasiperiodic breather dynamics is obscured from virtually all common diagnostics, including autocorrelation and other femtosecond pulse characterization techniques. Given the reduced pulse-to-pulse coherence in FM combs, time domain applications like the seeding of subsequent amplification stages or applications in ultrafast spectroscopy can probably be ruled out. With their surprisingly narrow beat notes, however, FM combs appear to be promising for frequency domain applications. In fact, given the miniature size of semiconductor lasers, there seems a lot of potential to replace bulky laser-based combs in applications like dual comb spectroscopy. However, such applications often heavily rely on spectral coherence, and this may not always be a given in broadband FM combs. In particular, we observed a clustering effect \cite{Cluster} in some of our simulations, i.e., the spectrum fragments into mutually incoherent groups of synchronized modes. Using our phase transition analogy, this clustering effect can be understood similar to the formation of droplets in a cloud, i.e., the condensation process localizes, and overarching coherence is lost. Consequently, the resulting combs are not equidistant anymore, which appears prohibitive for frequency domain applications of combs. However, if such clustering effects can be safely ruled out by suitable spectral coherence measurements, FM combs appear to be an appealing technology platform for a large variety of applications.

\bibliography{ChaoSync}

\end{document}